\documentclass[aps,showpacs,preprint,superscriptaddress]{revtex4}
\usepackage{graphicx}
\usepackage{subfigure}
\usepackage{float}
\usepackage{amsmath}
\usepackage{amsfonts}
\usepackage{color}
\usepackage{bm}
\usepackage{bbm}
\usepackage{txfonts}
\usepackage{array}
\usepackage{amssymb}

\begin{document}

\title{Chirp effects on pair production in oscillating electric fields with spatial inhomogeneity}
\author{Mamutjan Ababekri}
\affiliation{College of Mathematics and System Sciences, Xinjiang University, Urumqi 830046, China}
\author{Sayipjamal Dulat}
\affiliation{School of Physics Science and Technology, Xinjiang University, Urumqi 830046, China}
\author{B.S. Xie \footnote{bsxie@bnu.edu.cn}}
\affiliation{Key Laboratory of Beam Technology of the Ministry of Education, and College of Nuclear Science and Technology, Beijing Normal University, Beijing 100875, China}
\affiliation{Beijing Radiation Center, Beijing 100875, China}
\author{Jun Zhang \footnote{zhj@xju.edu.cn}}
\affiliation{School of Physics Science and Technology, Xinjiang University, Urumqi 830046, China}
\date{\today}

\begin{abstract}
Dirac-Heisenberg-Wigner formalism is used to study chirp effects on the vacuum pair creation under inhomogeneous electric fields.
For rapidly oscillating electric fields, the particle momentum spectrum is sensitive to both of the spatial scale and the chirp parameter, and the external field width has less significant effect for the maximally large chirp.
For slowly oscillating electric fields, chirp effects could be identified at large spatial extents and the carrier phase plays a significant role reflecting chirp effects even at small spatial scales.
We also notice that, the local density approximation holds for all external field profiles considered in this work at the quasihomogeneous limit allowing one to use arguments from homogeneous scenarios to analyze inhomogeneous results.
\end{abstract}
\maketitle

\section{Introduction}
Decay of the vacuum state under intense external fields is one of the nonperturbative predictions of the quantum electrodynamics(QED) yet to be observed \cite{Sauter:1931zz,Heisenberg:1935qt,Schwinger:1951nm}.
Reflecting the consequence of the quantum vacuum in the background field, detection of this nonlinear QED phenomena requires field strength of order $E_{\text{cr}} \sim 10^{16} \text{V/cm}$\cite{Dunne:2008kc}.
Although matter creation using strong external fields has been beyond the reach of experiments, electron positron creation in the laser experiments \cite{Burke:1997ew,Bamber:1999zt} and the realization of the light-by-light scattering of quasi-real photons in the particle accelerators \cite{dEnterria:2013zqi,Aaboud:2017bwk} have strengthened the interest in further quests for pair production studies.
The upcoming experiments play a major stimulus on this journey by promising laser intensities at the unprecedented level which makes it possible to observe nonperturbative vacuum pair production in the near future \cite{Ringwald:2001ib,Heinzl:2008an,Marklund:2008gj,Pike:2014wha}.

Theoretically, it is crucial to consider more realistic external field modes to provide reliable predictions for future experiments; for a review see Ref. \cite{Gelis:2015kya}.
The nonlinear nature of the vacuum pair creation causes the process to be sensitive to external field parameters, thus careful shaping of the applied field might induce special momentum signatures reflecting specific field structures and corresponding particle creation dynamics \cite{Hebenstreit:2009km,Dumlu:2010ua,Dumlu:2010vv,Dumlu:2011rr,Akkermans:2011yn}.
Moreover, the possibility of lowering the required field strength is reported for the cleaver combination of external fields with different frequencies \cite{Schutzhold:2008pz}.

Both of the characteristic momentum spectrum signatures and the enhancement mechanism are found in the pair creation under chirped laser field modes \cite{Dumlu:2010vv,Olugh:2018seh}.
Such frequency variations of the external field are crucial to be studied not only because they represent more complex field forms but for the fact that intense laser fields are realized in laboratories using the chirped pulse amplification(CPA) technique \cite{Strickland:1985gxr}.
The chirp effect is considered in tunneling and multiphoton absorption modes, and the effects of chirp parameters are mostly captured in the momentum spectrum, while the total particle number is raised drastically for large chirp parameters in the multiphoton process \cite{Dumlu:2010vv,Olugh:2018seh}.

On the other hand, previous works on pair creation studies in the inhomogeneous mode indicate that the spatial dependence of the external field may have nontrivial effects.
Particle selfbunching was reported in the momentum spectrum of particles in Schwinger pair creation where the total particle number also depends nonlinearly on the spatial scale of the external field \cite{Hebenstreit:2011wk}.
For an oscillating profile, the spatial focusing of the electric field introduces ponderomotive force effects into multiphoton process due to strong spatial variation of the electric field \cite{Kohlfurst:2017hbd}.
Moreover, a new kind of oscillatory pattern was noticed in Schwinger pair creation for the narrow spatial focusing \cite{Ababekri:2019dkl}.
These findings further indicate the importance of including spatial variations of the external field in pair production studies under more realistic field forms; see Ref. \cite{Aleksandrov:2017mtq} for further discussions.

In this paper, we use the real time Dirac-Heisenberg-Wigner(DHW) formalism \cite{Vasak:1987um,Hebenstreit:2011wk} to investigate chirp effects in the spatially inhomogeneous mode using simplified model for a chirped laser pulse.
A large span of spatial scales are considered to reflect the extent of spatial dependence of the external field.
In the quasihomogeneous scenario, where the spatial extent is large, we compare our results with pair creation in homogeneous cases.
When the spatial extent is small, we investigate results for differently oscillating external fields which may display various mechanisms due to the complex interplay between temporal and spatial parameters.
Special care is given to the carrier phase effect in the slowly oscillating scenario.

Our paper is organized as follows.
In Sec.~\ref{method}, we present the treatment of pair creation in 1+1 dimensions by introducing the field model to be considered in this study in Sec.~\ref{fields} and reviewing the key points of the DHW formalism in Sec.~\ref{DHWformalism}.
In Sec.~\ref{results}, the numerical results obtained for various filed shapes are presented with physical implications.
Sec.~\ref{result1} and Sec.~\ref{result2} show the momentum distribution as well as the particle yield for fast and slow oscillations respectively.
We present our conclusion in Sec.~\ref{summary}.

Throughout this article, natural units($ \hbar = c = 1 $) are used and the quantities are presented in terms of the electron mass $m$.

\section{Oscillating electric fields and pair creation}\label{method}

\subsection{External fields}\label{fields}
In this article, we study the electron positron pair production in 1+1 dimensions by considering the following oscillating electric field mode with space and time dependencies:
\begin{equation}\label{FieldMode}
\begin{aligned}
E\left(x,t\right)
&=E_{0} f \left( x \right ) g\left( t \right )\\
&=\epsilon \, E_{cr} \exp \left(-\frac{x^{2}}{2 \lambda^{2}} \right ) \exp \left(-\frac{t^{2}}{2 \tau^{2}} \right ) \cos(b t^2 + \omega t + \phi),
\end{aligned}
\end{equation}
where $E_{cr}$ is the critical field strength and $\lambda$ reflects the spatial scale of the external field.
We choose $E_{0} = 0.5 E_{\text{cr}}$($\epsilon=0.5$) and set $\omega=0.7 m$ with $\tau=45 m^{-1}$ when studying the rapidly oscillating mode, while for the slowly oscillating field, we let $\omega=0.1 m$ and  $\tau=25 m^{-1}$.
The introduction of a nonzero chirp $b$ results in a time dependent frequency which we call the effective frequency: $\omega_{\text{eff}}(t)=\omega+b t$.
Since pair production mostly occurs at the maximum of the external field and its nearby regions, we assume the relevant time interval to be $-\tau \le t \le \tau$(note that $\tau$ takes above chosen two values for corresponding oscillation modes) when estimating the range of the chirp parameter value.
We present the value for chirp in the form of $b= \alpha \omega/\tau$ ($\alpha \ge 0$)so that it indicates the change in the effective frequency by chirp in the relevant time interval $-\tau \le t \le \tau$:
\begin{equation}\label{chirp_def}
\omega_{\text{eff}}(\pm \tau)=\omega \pm b \tau = \omega \pm \alpha \omega.
\end{equation}
Then we set the theoretical upper limit for the chirp parameter by letting the maximum effective frequency to be around the threshold frequency $\omega_{\text{eff}}(\tau) \sim 1.0m$ for the fast pulse.
And for $\omega=0.1m$, the maximum chirp is taken with the condition $\omega_{\text{eff}}(\tau) \tau \sim \mathcal{O}(1)$ to preserve the fewcycle pulse shape of the external field.
Therefore, we take maximum chirp for $\omega=0.7 m$ to be $b=0.5 \omega/\tau \approx 0.0078 m^2$  and for $\omega=0.1 m$ the maximum chirp is $b=1.5\omega/\tau = 0.006 m^2$.

In this idealized standing wave profile, we attempt to model the oscillating electric field with spatial dependency constructed by two counter propagating coherent laser fields aligned to cancel out the magnetic component.  
We point out that the direction of the field is along the $x$-axis and field strength varies in both $x$ and $t$.
Through this simplified model, we are investigating how spatial variation and other temporal field parameters interplay with each other and affect pair creation by calculating the produced particles' number density in the phase space.

\subsection{DHW formalism}\label{DHWformalism}
The DHW method was developed from the fermion density operator to explore the phase space structure of the Dirac vacuum \cite{Vasak:1987um}, and it is an efficient formalism for the investigations on pair creation under both spatially homogeneous \cite{Hebenstreit:2010vz,Blinne:2013via,Li:2015cea,Olugh:2018seh,Olugh:2019nej} as well as inhomogeneous electromagnetic fields \cite{Hebenstreit:2011wk,Kohlfurst:2017hbd,Kohlfurst:2015niu,Ababekri:2019dkl}.
In this paper, the DHW equations of motion are solved numerically \cite{Kohlfurst:2015zxi,Ababekri:2019dkl} for spatially inhomogeneous external field modes given in Eq.\eqref{FieldMode}.
Since the detailed calculations and numerical strategies could be found in Refs.~\cite{Hebenstreit:2011wk,Kohlfurst:2015zxi,Kohlfurst:2017hbd,Ababekri:2019dkl}, we only provide the equations and observable quantities relevant to our study in the following.

The complete set of equations of motion for the Wigner components are reduced to 4 equations in the 1+1 dimensional scenario \cite{Hebenstreit:2011wk,Kohlfurst:2017hbd,Ababekri:2019dkl}:
\begin{align}
 &D_t \mathbbm{s} - 2 p_x \mathbbm{p} = 0 , \label{pde:1}\\
 &D_t \mathbbm{v}_{0} + \partial _{x} \mathbbm{v} = 0 , \label{pde:2}\\
 &D_t \mathbbm{v} + \partial _{x} \mathbbm{v}_{0} = -2 m \mathbbm{p} , \label{pde:3}\\
 &D_t \mathbbm{p} + 2 p_x \mathbbm{s} = 2 m \mathbbm{v} , \label{pde:4}
\end{align}
with the pseudodifferential operator
\begin{equation}\label{pseudoDiff}
 D_t = \partial_{t} + e \int_{-1/2}^{1/2} d \xi \,\,\, E_{x} \left( x + i \xi \partial_{p_{x}} \, , t \right) \partial_{p_{x}} .
\end{equation}
Among the four Wigner components in the the equations of motion\eqref{pde:1}-\eqref{pde:4} we have only two non-vanishing vacuum initial conditions \cite{Kohlfurst:2015zxi}:
\begin{equation}\label{vacuum-initial}
{\mathbbm s}_{vac} = - \frac{2m}{\omega} \, ,
\quad  {\mathbbm v}_{vac} = - \frac{2{ p_x} }{\omega} \,  ,
\end{equation}
where $\omega=\sqrt{p_{x}^{2}+m^2}$ is the energy of a particle.
By explicitly subtracting these vacuum terms, the modified Wigner components are written as:
\begin{equation}\label{vacuum-initial-}
{\mathbbm w}^{v} = {\mathbbm w} - {\mathbbm w}_{vac},
\end{equation}
where ${\mathbbm w}$ denotes the four Wigner components in our 1+1 formalism.
Then the particle number density in the phase space could be defined via dividing the total energy of the created particles by individual particle energy \cite{Hebenstreit:2011wk}:
\begin{equation}\label{PS}
n \left( x , p_{x} , t \right) = \frac{m  \mathbbm{s}^{v} \left( x , p_{x} , t \right) + p_{x}  \mathbbm{v}^{v} \left( x , p_{x} , t \right)}{\omega \left( p_{x} \right)}.
\end{equation}
The position distribution or the momentum distribution of the created particles could be obtained  from $n \left( x , p_{x} , t \right)$  by integrating out $p_x$ or $x$ respectively, and the total particle yield is calculated by integrating over the whole phase space:
\begin{equation}\label{Num}
N\left(t \right) = \int dx dp_x n \left( x , p_{x} , t \right).
\end{equation}

The numerical treatment for Eqs. \eqref{pde:1}-\eqref{pde:4} follows from the techniques developed in Ref. \cite{Kohlfurst:2015zxi}, where the spectral method is employed to handle the pseudodifferential operator \eqref{pseudoDiff}.
The calculation parameters are chosen in similar way as in Refs. \cite{Kohlfurst:2015zxi,Ababekri:2019dkl} and tested for reasonable parameter space such that final results are convergent.

\section{Numerical results}\label{results}

\subsection{Rapidly oscillating electric field: $\omega = 0.7 m$.}\label{result1}

In this subsection, we obtain results for $\omega=0.7 m$ characterizing pair production in the rapidly oscillating electric fields with spatial focusing using external field model \eqref{FieldMode} for various chirp parameter values.
We set $\tau=45 m^{-1}$ such that, for $b=0$(see Fig. \ref{omg07tau45b0ph0}), we could obtain same results as in Fig. 3 of Ref. \cite{Kohlfurst:2017hbd}, where the temporal pulse envelope is chosen as $\cos^4(\frac{t}{\tau})$ with $\tau=100 m^{-1}$.
Before we discuss the results for nonzero chirp values, we investigate in detail the pair production process for $b=0$ so that it may assist the discussions for the rest of the calculations.

\subsubsection{Local density approximation.}\label{result1a}
\begin{figure}[ht]\suppressfloats
\includegraphics[scale=0.6]{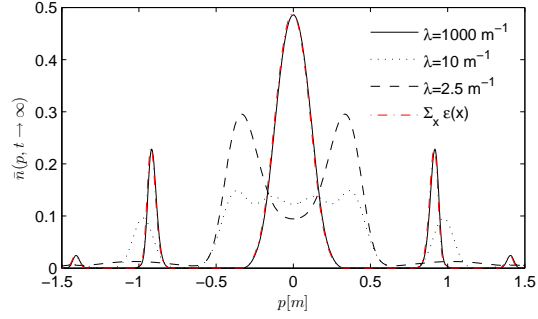}
\caption{Reduced momentum spectrum for spatially focused oscillating electric fields \eqref{FieldMode} with $\omega = 0.7 m$ and $b=0$.  Other field parameters are $\epsilon=0.5$, $\tau=45 m^{-1}$ and $\phi=0$. Note that, the summed momentum by the local density approximation labeled as $\Sigma_{x}\epsilon(x)$ (red dot-dashed curve) overlaps with the quasihomogeneous result.  }
\label{omg07tau45b0ph0}
\end{figure}

In Fig. \ref{omg07tau45b0ph0}, we recover the signatures of multiphoton absorbtion of the homogeneous scenario in the quasihomogeneous limit where $\lambda=1000 m^{-1}$.
The n-photon absorption spectrum could be understood as a result of particle trajectory adding up to form interference patterns\cite{Nousch:2015pja}.
When the spatial width of the external field decrease to $\lambda=10 m^{-1}$, the main peak at $p=0$ in the momentum spectrum spreads and displays an oscillatory pattern due to the disruption of a coherent superposition of particle trajectories by the finite size of the pulse.
For $\lambda=2.5 m^{-1}$, the ponderomotive force caused by the highly inhomogeneous nature of the external field accelerates particles in the directions where the external field decrease resulting in the split of the $p=0$ peak\cite{Kohlfurst:2017hbd}.

\begin{figure}[ht]\suppressfloats
\includegraphics[scale=0.6]{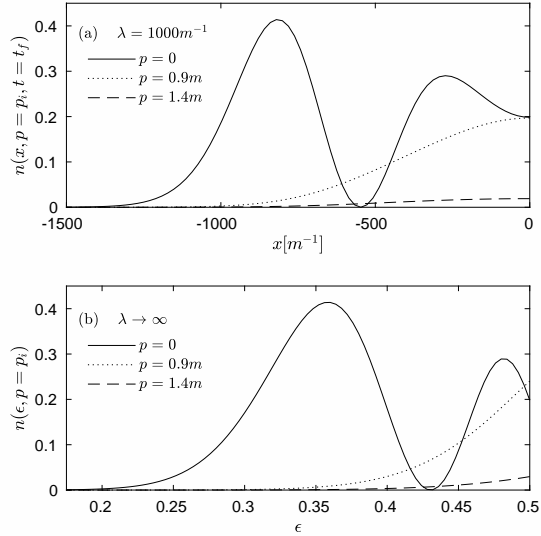}
\caption{The position distribution for typical momentum peaks $n(x,p=p_i,t=t_f)$ for $\lambda=1000 m^{-1}$(top) and corresponding peak values $n(\epsilon, p=p_i)$ for various field strengths $\epsilon$ for homogeneous field (bottom).  The asymptotic final time is set to $t_{f}=6.5 \tau$.  }
\label{omg07tau45b0ph0_pi}
\end{figure}

However, compared with the homogenous limit($\lambda \rightarrow \infty$), the main peak at vanishing momentum seems to be more pronounced in the quasihomogeneous scenario as is noticed in Ref. \cite{Kohlfurst:2017hbd}.
Furthermore, from the position distribution of certain momenta $n(x,p=p_i,t=t_f)$ shown in Fig. \ref{omg07tau45b0ph0_pi}(a), we observe that, with the departure of particle location from the origin, the heights of typical momentum peaks behave like that of pair production under homogeneous fields with decreasing field strength $\epsilon$:
The related peak values would decrease for $p=0.9 m$ and $p=1.4 m$, while the vanishing momentum peak would reach maximums and minimum with decreasing $\epsilon$, see Fig. \ref{omg07tau45b0ph0_pi}(b).
It seems that the spatial dependency of the external field is playing the role of an effective field strength such that at different locations the field obtains different amplitudes according to the gaussian function of the position $\epsilon(x) = 0.5 \exp(-\frac{x^2}{2 \lambda^2})$, and the pair production process occurs independently at each location.

By this observation, we could calculate a new momentum distribution by summing results for homogeneous fields with different field strengths given as:
\begin{equation}\label{effDis}
\tilde{n}(p, t \rightarrow \infty) = \sum_{x} \frac{n(\epsilon(x)| p, t \rightarrow \infty)}{\lambda},
\end{equation}
where $n(\epsilon(x)| p, t \rightarrow \infty)$ is the momentum distribution for the homogeneous field $E(t)$ with effective field strengthes:
\begin{equation}
E(t) = \epsilon(x) E_{cr} \exp \left(-\frac{t^{2}}{2 \tau^{2}} \right ) \cos(b t^2 + \omega t + \phi).
\end{equation}
And the spatial scale $\lambda$ in the denominator in Eq.~\eqref{effDis} is introduced since we calculate the reduced particle number distribution.
The corresponding result agrees with the quasihomogeneous calculation as shown in Fig. \ref{omg07tau45b0ph0} where the red dot-dashed curve is the approximation result.
Thus we may understand the large maximum peak around $p=0$ for $\lambda=1000 m^{-1}$ in Fig. \ref{omg07tau45b0ph0} by referring to pair creation in homogeneous fields where the maximally large peak value is present for smaller effective strengthes which is included in the summation \eqref{effDis}, see Fig. \ref{omg07tau45b0ph0_pi}(b).
Nevertheless, we could relate the overall shape of the momentum spectrum for $\lambda = 1000 m^{-1}$ to pair creation for homogeneous field with same $\epsilon=0.5$.

This local density approximation(LDA) that we have used above for a multiphoton absorption process is only valid when the pair-formation length $l=2m/|e \epsilon E_{cr}|$ is much smaller than the spatial width $\lambda$ of the external field\cite{Hebenstreit:2011wk,Aleksandrov:2018zso,Aleksandrov:2019ddt}.   We have noticed that, not only the summed momentum of the LDA calculation agrees with the exact result for an inhomogeneous field(see the red line and the black solid line in Fig. \ref{omg07tau45b0ph0}), the typical momentum peak at each location $n(x, p_i)$ also closely related to the homogeneous result with $\epsilon(x)$at each location $x$(see Figs. \ref{omg07tau45b0ph0_pi}(a) and (b)).
This profound relation between details of two approaches is due to the formation of particle pairs at each location independently via photon absorption.
This detailed relation is weaken in a tunneling process, however, it is still sufficient to find agreement between the summed momentum(the LDA calculation) and the quasihomogeneous results so that momentum distribution could be explained in terms of temporal pulse structure.

\subsubsection{Chirp effects.}\label{result1b}

\begin{figure}[ht]\suppressfloats
\includegraphics[scale=0.5]{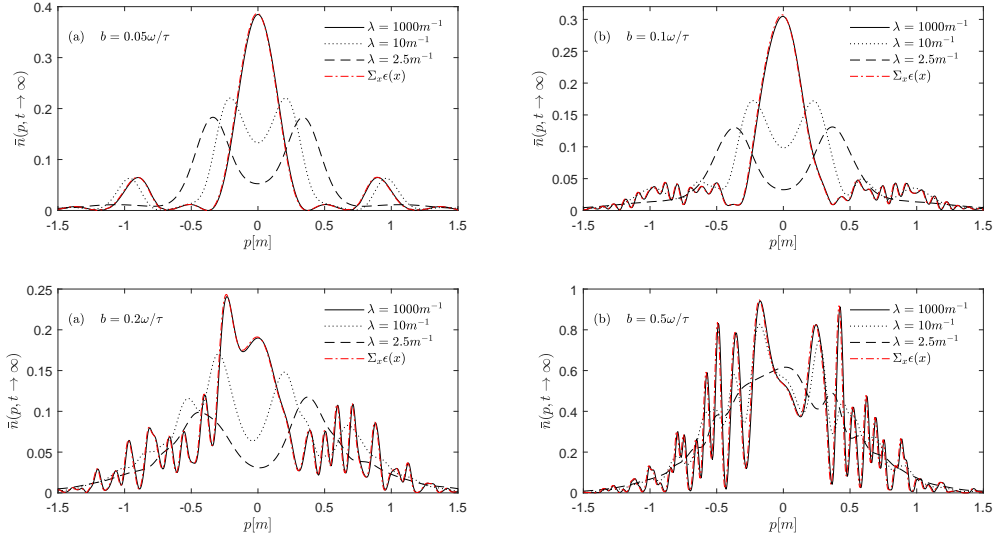}
\caption{Reduced momentum spectrum for spatially focused oscillating electric fields \eqref{FieldMode} with nonzero chirp values. Other field parameters are same as the zero chirp case in Fig. \ref{omg07tau45b0ph0}(a). The corresponding increasing chirp values are $b \approx 0.00078 m^2, 0.0016 m^2, 0.0031 m^2$ and $0.0078 m^2$. }
\label{omg07tau45b1b2b3b4ph0}
\end{figure}

If we assume the dominant contribution from external field to pair creation comes in the time interval from $-\tau$ to $\tau$, the change in the time dependent frequency $\omega_{\text{eff}}=\omega + b t$ by the linear chirp $b$ could be estimated as:
\begin{equation}\label{deltaomega}
\Delta \omega_{\text{eff}} = \omega_{\text{eff}}(\tau) - \omega_{\text{eff}}(-\tau).
\end{equation}
For the chirp values considered in Fig. \ref{omg07tau45b1b2b3b4ph0}, $\Delta \omega$ ranges from $0.1 \omega$ to $1.0 \omega$ and the corresponding momentum distribution differs greatly from the $b=0$ case in Fig. \ref{omg07tau45b0ph0} for larger chirp.

For $\lambda=1000 m^{-1}$, the momentum distribution could be directly related to the temporal pulse structure.
The disruption of the constant frequency by the chirp weakens the n-photon absorption spectrum in Figs. \ref{omg07tau45b1b2b3b4ph0}(a) and (b).
However, since $\Delta\omega_{\text{eff}} \le 0.2 \omega$ for these small chirp values, we could still observe the main peak at $p=0$ which corresponds to the 3-photon absorption in the zero chirp case.
For large chirp values, the resulting momentum distribution completely differs from the multiphoton spectrum displaying a complex oscillation, see Figs. \ref{omg07tau45b1b2b3b4ph0}(c) and (d).
Since pair creation is a non-Markovian process where earlier history of the events are accumulated into the process\cite{Schmidt:1998zh}, wide range of change in the time dependent frequency $\omega_{\text{eff}}(t)$ has significant influences.

When spatial focusing size decreases to $\lambda=10 m^{-1}$, both temporal structure and the finite spatial width become crucial.
The finite size of the external field prevents particle formation to be dominated by the temporal pulse structure.
However, we could observe momentum spectrum displaying the same complicated oscillating pattern as in the quasihomogeneous case for large chirp parameters indicating that finite spatial width plays less significant role for such large chirp, see Fig. \ref{omg07tau45b1b2b3b4ph0}(d).

At the extremely small spatial width $\lambda=2.5 m^{-1}$, the peak splitting present in $b=0$ is present till the chirp parameter reaches $b=0.5 \omega/\tau$ where we loose such effect in the momentum spectrum.
This is related to the highly nonuniform oscillation caused by the large chirp parameter inhibiting the formation for ponderomotive force which pushes particles towards low field intensity regions in space~\cite{Kohlfurst:2017hbd}.

\begin{figure}[ht]\suppressfloats
\includegraphics[scale=0.6]{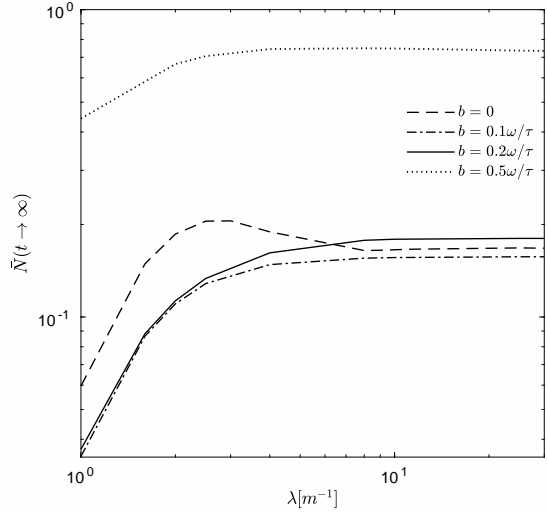}
\caption{Reduced particle yield at various spatial extents for chirped oscillating electric fields \eqref{FieldMode} with frequency $\omega=0.7 m$.
The dotted line for the maximal chirp $b=0.5 \omega/\tau$ shows the overall tendency of the total particle numbers with the increasing chirp at different spatial scales.  }
\label{N_omg07}
\end{figure}

The effects of chirp parameter at various spatial extents could also be seen in the total particle yield presented in Fig. \ref{N_omg07}.
As is reported in Ref.~\cite{Kohlfurst:2017hbd}, a possible dynamical enhancement present for the $\omega=0.7 m$ oscillation mode, and this enhancement seems to benefit from the strong ponderomotive force at narrow spatial scale around $\lambda \sim 2 m^{-1}$.
The decrease of particle yield for $b=0.1 \omega/\tau$ indicates that small chirp changes the oscillation frequency and restrains the enhancement mechanism\cite{Linder:2015vta}.
With the further increase of chirp value, higher effective frequencies at later times $\omega_{\text{eff}}(t)$ dominate the process and increases the total yield.
For the maximal chirp $b=0.5 \omega/\tau$, particles are created mostly by high energy photon absorption such that particle number increases greatly and the spatial dependency of pair creation would become less significant.

\subsection{Slowly oscillating electric field: $\omega = 0.1 m$. }\label{result2}

Now we consider the slowly oscillating electric field mode by choosing $\omega=0.1 m$ and split the results for $\phi=0$ and $\phi=\pi/2$ since envelope phase has crucial effect on the temporal pulse shape.
The temporal pulse length is chosen as $\tau= 25 m^{-1}$, so that we achieve numerical convenience while retaining the main features of pair creation in electric fields with subcycle structure \cite{Hebenstreit:2009km,Dumlu:2010vv}.
In this tunneling dominated regime, we have also recovered the quasihomogeneous results by applying the LDA calculations, see red dot-dashed lines shown in Fig. \ref{omg01b0b1b2b3ph0} and Fig. \ref{omg01b0b1b2b3ph2}.

\subsubsection{$\phi=0$}\label{result2a}

\begin{figure}[ht]\suppressfloats
\includegraphics[scale=0.5]{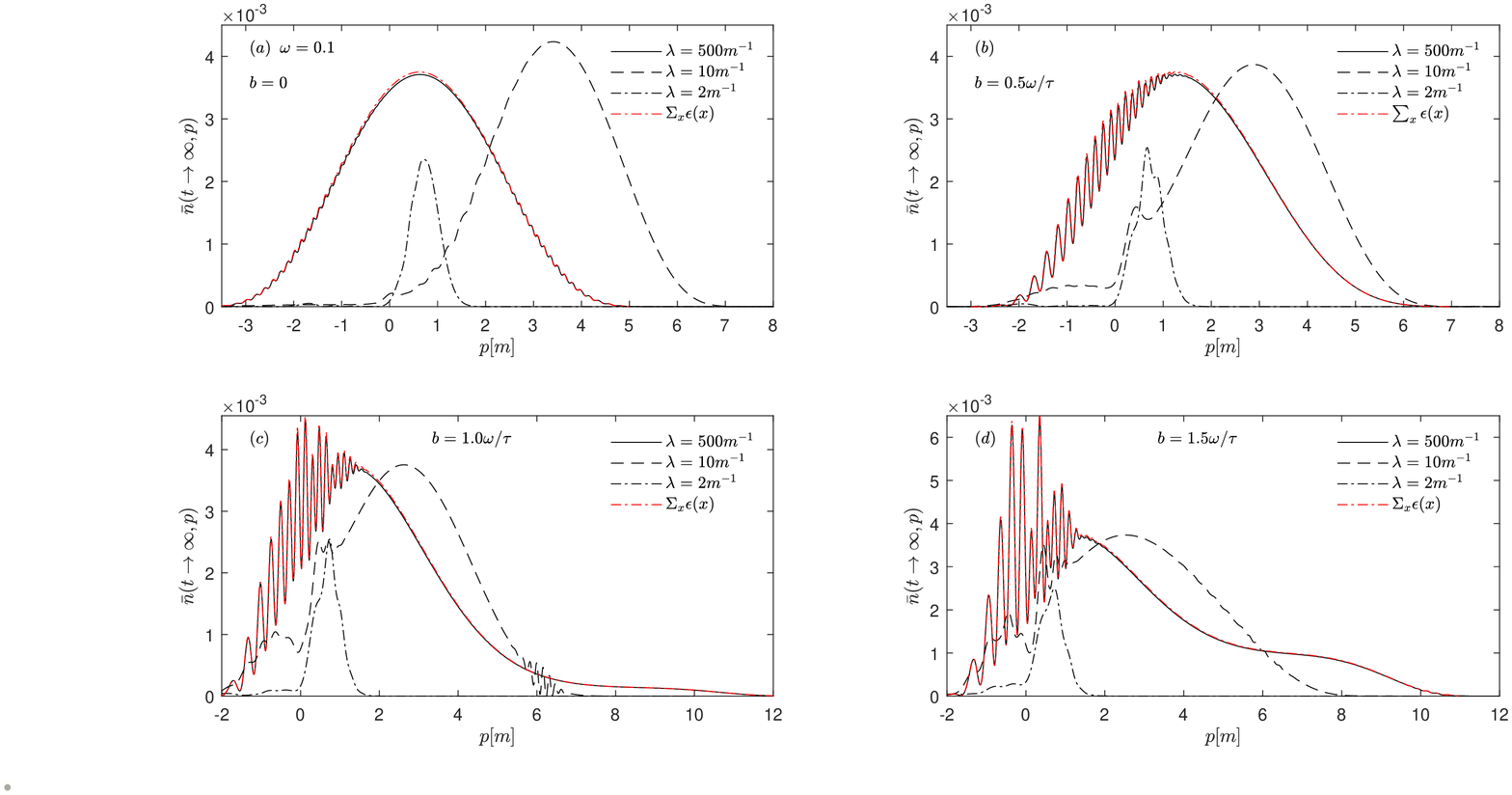}
\caption{Reduced momentum spectrum at various spatial extents for chirped oscillating electric fields \eqref{FieldMode} with frequency $\omega=0.1 m$, $\tau=25 m^{-1}$ and $\phi=0$. The corresponding increasing chirp values are $b = 0, 0.002 m^2, 0.004 m^2$ and $0.006 m^2$. }
\label{omg01b0b1b2b3ph0}
\end{figure}

Fig. \ref{omg01b0b1b2b3ph0}(a) shows the momentum distribution for pair creation under electric fields with subcycle structures at various spatial scales.
At $\lambda=500 m^{-1}$, the momentum distribution is similar to the homogenous case and does not center around $p=0$ because of the small oscillation frequency.
When spatial width decreases to $\lambda=10 m^{-1}$ momentum spectrum shifts towards higher momentum values.
This shift in the momentum distribution is different from the case for the Schwinger pair creation in a single pulse $\text{sech}^2(t/\tau)$ \cite{Hebenstreit:2011wk}.
In the single pulse field, decrease in spatial width causes particle selfbunching where momentum spectrum shifts towards the vanishing momentum direction with higher peak value because particles leave the finite field region and can not be accelerated.
In the present case the width of electric field is finite, particles created with certain momentum leave the electric filed region and miss the particle deceleration by the negative field peak.
Also particles leaving the external field region could miss the chance to interfere with other particles such that the oscillatory
pattern present for $\lambda = 500 m^{-1}$ is weaken or missing.
Therefore, lack of particle-field interactions and lack of interference with other particles affects the momentum distribution and may cause a shift in the particle momentum distribution.
\begin{figure}[ht]\suppressfloats
\includegraphics[scale=0.75]{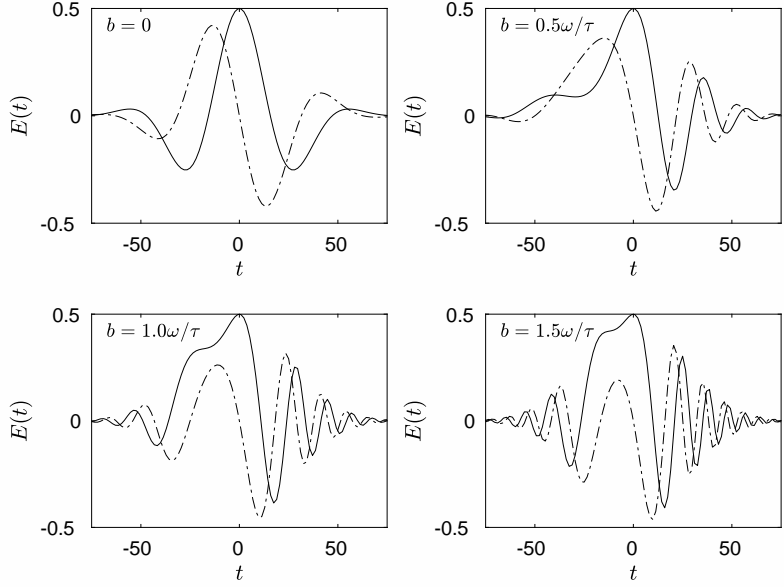}
\caption{Temporal electric field shape for $E(t)=0.5 \exp(-\frac{t}{2 \tau^2})\cos(\omega t + b t^2)$ with $\omega=0.1$ and $\tau=25$. }
\label{temporalfield}
\end{figure}

At the quasihomogeneous case of $\lambda=500 m^{-1}$, the small chirp parameter causes stronger oscillatory pattern in the momentum spectrum compared with the $b=0$ case, see Figs. \ref{omg01b0b1b2b3ph0}(a) and (b).
This could be understood as the interference of particles created from the maximum main peak and the opposite signed side peak of the temporal field\cite{Hebenstreit:2009km}, see Fig. \ref{temporalfield} for the temporal pulse shape.
Momentum distribution range widens and shifts especially for larger chirp values in Figs. \ref{omg01b0b1b2b3ph0}(c) and (d) because external field oscillation slows down greatly within the dominant time interval and causes particle acceleration.
For instance, when chirp $b=1.5 \omega/\tau$, we have,
\begin{equation}\label{zerofrequency}
\omega_{\text{eff}}(t=-\frac{2}{3}\tau) = \omega + b t = 0.
\end{equation}
Therefor, oscillating field peaks with vanishing frequency or wider temporal width appear within $-\tau \le t \le \tau$ and spreads momentum distribution range.

At the strong focused scenario, where $\lambda$ is small, particles leave from the external field region and chirp effects, such as strong oscillations and shift in the spectrum, are not as pronounced as in the large spatial extent case.

\subsubsection{$\phi=\frac{\pi}{2}$}\label{result2b}

\begin{figure}[ht]\suppressfloats
\includegraphics[scale=0.5]{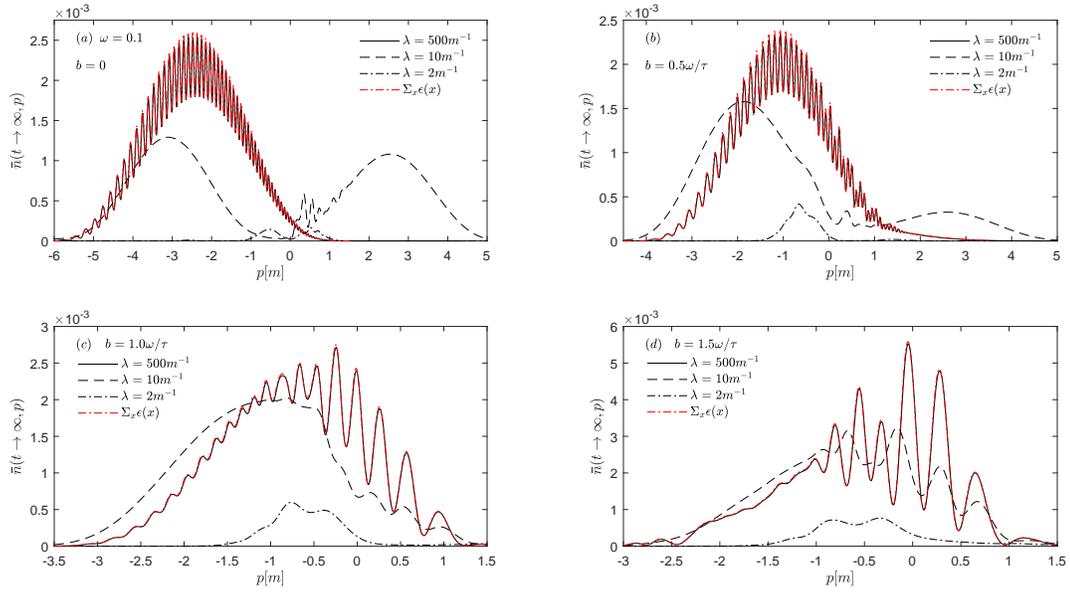}
\caption{Reduced momentum spectrum at at various spatial extents for chirped oscillating electric fields \eqref{FieldMode} with frequency $\omega=0.1 m$, $\tau=25 m^{-1}$ and $\phi=\pi/2$. The corresponding increasing chirp values are $b =0, 0.002 m^2, 0.004 m^2$ and $0.006 m^2$. Note that, chirp effect is noticeable even for $\lambda=10 m^{-1}$. }
\label{omg01b0b1b2b3ph2}
\end{figure}

For $\phi=\pi/2$ and $\lambda=500 m^{-1}$, the nonzero chirp changes the coherent interference pattern in Fig. \ref{omg01b0b1b2b3ph2}(a) and turns it into a complex oscillatory one for the large chirp shown in Fig. \ref{omg01b0b1b2b3ph2}(d).
The interference pattern for $b=0$ could be understood as the interference between temporally separated pair creation events \cite{Hebenstreit:2009km} and the complex oscillation for large chirp as the consequence of pair creation events from multiple sources, see the dot-dashed line in Fig. \ref{temporalfield} for the temporal pulse shape.
Compared with the homogeneous case, where asymptotic minimum in the momentum spectrum would reach the bottom, the quasihomogeneous result gives a weaker interference pattern.
This weakening effect could be understood in the light of the LDA where some of the complete interference patterns for different effective field strengthes cancel out in the sum.

Interestingly, compared with the results for $\phi=0$, the chirp affects momentum spectrum differently for $\lambda=10 m^{-1}$ in Fig. \ref{omg01b0b1b2b3ph2}.
Particles created from two opposite sign field peaks with large temporal durations leave the finite field region in two directions and form two momentum peaks for $\lambda=10 m^{-1}$ in Fig. \ref{omg01b0b1b2b3ph2}(a).
With the increase of $b$, relatively fast oscillating peaks with shorter temporal duration in the external field is less affected by the finite spatial width and we observe the merging of two major momentum peaks finally forming an oscillatory structure similar to the complex oscillation in the quasihomogeneous case, see Fig. \ref{omg01b0b1b2b3ph2}(d).

\begin{figure}[ht]\suppressfloats
\includegraphics[scale=0.6]{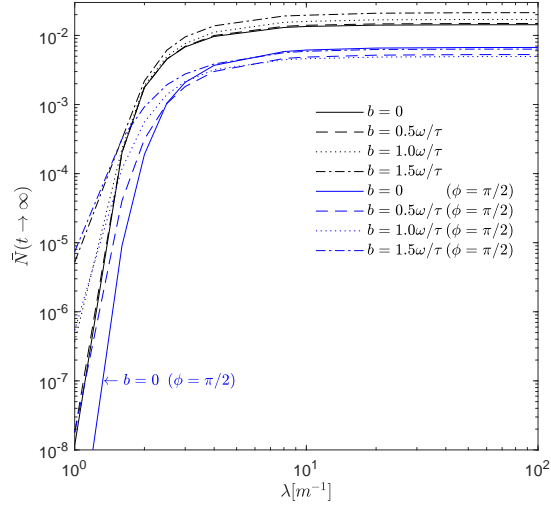}
\caption{Reduced particle yield at various spatial extents for chirped oscillating electric fields \eqref{FieldMode} with frequency $\omega=0.1 m$.
The blue lines, which are below the black lines at the large spatial scales, correspond to $\phi=\pi/2$ results.  }
\label{N1-2}
\end{figure}

In Fig. \ref{N1-2}, we display the reduced total numbers of created particles for various spatial extents.
At large spatial scales, final particle yield is smaller for $\phi=\pi/2$ because of the decrease of the external field energy by changing from the cosine field with maximal peak at $t=0$ into two smaller side peaks of a sine field, see Fig. \ref{temporalfield}.
However, the chirp parameter seems to affect the total particle yield differently for different phase values which is also explained by the changes in the temporal field shape.
Around the electron Compton wavelength, sudden drop in the total yield reflects the general feature for tunneling mode where external field's total energy becomes too small to produce particles\cite{Hebenstreit:2011wk}.
For narrow spatial scales, the total number is sensitive to both chirp and carrier envelope phase parameters.
At these cases pair creation is no longer dominated by temporal field parameters, rather, it is a combined effect of both of the spatial scale and the temporal field structure calling for more general means to explain.

\section{Conclusion}\label{summary}
We have shown the chirp effects at various spatial scales of the external field for two oscillating frequencies.
When the spatial scale $\lambda \sim 1000 m^{-1}$, the external field form $E(t,x)=E_0 \exp(-\frac{x^2}{2 \lambda^2}) g(t)$ could be considered as the generalization of the homogeneous form $E(t)=E_0 g(t)$ such that two field forms have direct connection via local density approximation.
Thus the pair production features in the momentum spectrum could be analyzed using similar arguments from the homogeneous case.
Moreover, in the multiphoton absorption scenario, we notice the one to one correspondence between the particle number for certain momentum at individual positions $n(x_i,p)$ for the quasihomogeneous field and the homogenous results $n(\epsilon(x_i),p)$ with different field strengths .
This profoundness of the local density approximation maybe is to the adiabatic nature of multiphoton absorbtion process where particle creation happens by photon absorption instead of the work done by the external field.

At the large spatial scale, the small chirp parameter disrupts the n-photon absorption spectrum and large chirp results in a complex oscillatory distribution loosing the sign of a multiphoton absorption for rapidly oscillating field.
Large chirp also causes enhancement in the total particle number and is less affected by the spatial width because of the high frequency modes present in the time dependent frequency $\omega_{\text{eff}}(t)$.
For the slowly oscillating field, chirp causes shift and interference patterns in the momentum distribution.
Nonzero chirp causes slight change in total particle yield because of the change in the temporal field shape.

The strong spatial focusing reflects chirp effects differently for various oscillating modes.
When the electric field oscillates slowly, the temporal carrier phase effect is found to be crucial even for $\lambda=10 m^{-1}$.
The chirp effect is not obvious for $\phi=0$ at this scale, however, for $\phi=\pi/2$ we see merging and the formation of oscillatory pattern in the momentum spectrum with the increase of chirp value which is quite different from the homogeneous instance.
This indicates that the finite spatial scale renders effects of the temporal pulse structure differently according to the temporal pulse structure and complicates the nonMarkovian nature of the process \cite{Schmidt:1998zh} with the additional dimension.

These results also suggest that introducing finite spatial scales of the external fields may have crucial consequences to the homogenous results, thus one needs to be more cautious when calculating multidimensional external field results to provide more accurate predictions.
Also, the shape of the spatial pulse may have nontrivial effects on particle momentum spectrum and even on the total number enhancement.
Furthermore, analytic tools needs to be generalized to broader parameter ranges so that it is possible to better understand nontrivial features of pair creation in inhomogeneous fields.

\begin{acknowledgments}
\noindent
We thank Obulkasim Olugh for useful discussion and valuable comments.
We are also grateful to Nureli Yasen and Binbing Wu for their help with the numerical calculation.
And we are thankful to the anonymous referee for helpful comments and suggestions.
This work was supported by the National Natural Science Foundation of China (NSFC) under
Grant No.\ 11935008, 11875007 and 11864039.
The computation was carried out at the HSCC of the Beijing Normal University.

\end{acknowledgments}


\begin{thebibliography}{99}\suppressfloats

\bibitem{Sauter:1931zz}
  F.~Sauter,
  Z.\ Phys.\  {\bf 69}, 742 (1931).

\bibitem{Heisenberg:1935qt}
  W.~Heisenberg and H.~Euler,
  Z.\ Phys.\  {\bf 98}, 714 (1936).%

\bibitem{Schwinger:1951nm}
  J.~S.~Schwinger,
  Phys.\ Rev.\  {\bf 82}, 664 (1951).

\bibitem{Dunne:2008kc}
  G.~V.~Dunne,
  Eur.\ Phys.\ J.\ D {\bf 55}, 327 (2009).%

\bibitem{Burke:1997ew}
  D.~L.~Burke {\it et al.},
  Phys.\ Rev.\ Lett.\  {\bf 79}, 1626 (1997).

\bibitem{Bamber:1999zt}
  C.~Bamber {\it et al.},
  Phys.\ Rev.\ D {\bf 60}, 092004 (1999).

\bibitem{dEnterria:2013zqi}
  D.~d'Enterria and G.~G.~da Silveira,
  Phys.\ Rev.\ Lett.\  {\bf 111}, 080405 (2013);\\
  Erratum: [Phys.\ Rev.\ Lett.\  {\bf 116}, no. 12, 129901 (2016)].

\bibitem{Aaboud:2017bwk}
  M.~Aaboud {\it et al.} [ATLAS Collaboration],
  Nature Phys.\  {\bf 13}, no. 9, 852 (2017).

\bibitem{Ringwald:2001ib}
  A.~Ringwald,
  Phys.\ Lett.\ B {\bf 510}, 107 (2001).%

\bibitem{Heinzl:2008an}
  T.~Heinzl and A.~Ilderton,
  Eur.\ Phys.\ J.\ D {\bf 55}, 359 (2009).%

\bibitem{Marklund:2008gj}
   M.~Marklund and J.~Lundin,
  Eur.\ Phys.\ J.\ D {\bf 55}, 319 (2009).%

\bibitem{Pike:2014wha}
  O.~J.~Pike, F.~Mackenroth, E.~G.~Hill and S.~J.~Rose,
  Nature Photon.\  {\bf 8}, 434 (2014).

\bibitem{Gelis:2015kya}
  F.~Gelis and N.~Tanji,
  Prog.\ Part.\ Nucl.\ Phys.\  {\bf 87}, 1 (2016).%

\bibitem{Hebenstreit:2009km}
  F.~Hebenstreit, R.~Alkofer, G.~V.~Dunne and H.~Gies,
  Phys.\ Rev.\ Lett.\  {\bf 102}, 150404 (2009).%

\bibitem{Dumlu:2010vv}
  C.~K.~Dumlu,
  Phys.\ Rev.\ D {\bf 82}, 045007 (2010).%


\bibitem{Akkermans:2011yn}
  E.~Akkermans and G.~V.~Dunne,
  Phys.\ Rev.\ Lett.\  {\bf 108}, 030401 (2012).%

\bibitem{Dumlu:2010ua}
  C.~K.~Dumlu and G.~V.~Dunne,
  Phys.\ Rev.\ Lett.\  {\bf 104}, 250402 (2010).%

\bibitem{Dumlu:2011rr}
  C.~K.~Dumlu and G.~V.~Dunne,
  Phys.\ Rev.\ D {\bf 83}, 065028 (2011).%

\bibitem{Schutzhold:2008pz}
  R.~Schutzhold, H.~Gies and G.~Dunne,
  Phys.\ Rev.\ Lett.\  {\bf 101}, 130404 (2008).%

\bibitem{Olugh:2018seh}
  O.~Olugh, Z.~L.~Li, B.~S.~Xie and R.~Alkofer,
  Phys.\ Rev.\ D {\bf 99}, no. 3, 036003 (2019).%

\bibitem{Strickland:1985gxr}
  D.~Strickland and G.~Mourou,
  Opt.\ Commun.\  {\bf 55}, no. 6, 447 (1985); \\
  Erratum: [Opt.\ Commun.\  {\bf 56}, 219 (1985)].

\bibitem{Vasak:1987um}
  D.~Vasak, M.~Gyulassy and H.~T.~Elze,
  Annals Phys.\  {\bf 173} (1987) 462;\\
  I. Bialynicki-Birula, P. G\'ornicki and J. Rafelski,
  Phys. Rev. D {\bf 44} (1991) 1825; \\
  F.~Hebenstreit, R.~Alkofer and H.~Gies,
  Phys.\ Rev.\ D {\bf 82}, 105026 (2010).%

\bibitem{Hebenstreit:2011wk}
  F.~Hebenstreit, R.~Alkofer and H.~Gies,
  Phys.\ Rev.\ Lett.\  {\bf 107}, 180403 (2011).%

\bibitem{Kohlfurst:2017hbd}
  C.~Kohlf\"urst and R.~Alkofer,
  Phys.\ Rev.\ D {\bf 97}, no. 3, 036026 (2018).%

\bibitem{Ababekri:2019dkl}
  M.~Ababekri, B.~S.~Xie and J.~Zhang,
  Phys.\ Rev.\ D {\bf 100}, no. 1, 016003 (2019).

\bibitem{Aleksandrov:2017mtq}
  I.~A.~Aleksandrov, G.~Plunien and V.~M.~Shabaev,
  Phys.\ Rev.\ D {\bf 96}, no. 7, 076006 (2017);\\%
  I.~A.~Aleksandrov, G.~Plunien and V.~M.~Shabaev,
  Phys.\ Rev.\ D {\bf 97}, no. 11, 116001 (2018).%

\bibitem{Kohlfurst:2015niu}
  C.~Kohlf\"urst and R.~Alkofer,
  Phys.\ Lett.\ B {\bf 756}, 371 (2016);\\%
  C.~Kohlf\"urst,
  Eur.\ Phys.\ J.\ Plus {\bf 133}, no. 5, 191 (2018).%

\bibitem{Hebenstreit:2010vz}
  F.~Hebenstreit, R.~Alkofer and H.~Gies,
  Phys.\ Rev.\ D {\bf 82}, 105026 (2010).

\bibitem{Blinne:2013via}
  A.~Blinne and H.~Gies,
  Phys.\ Rev.\ D {\bf 89}, 085001 (2014);\\%
  A.~Blinne and E.~Strobel,
  Phys.\ Rev.\ D {\bf 93}, 025014 (2016).%

\bibitem{Li:2015cea}
  Z.~L.~Li, D.~Lu and B.~S.~Xie,
  Phys.\ Rev.\ D {\bf 92}, no. 8, 085001 (2015);\\%
  Z.~L.~Li, Y.~J.~Li and B.~S.~Xie,
  Phys.\ Rev.\ D {\bf 96}, 076010 (2017);\\%
  Z.~L.~Li, B.~S.~Xie and Y.~J.~Li,
  Phys.\ Rev.\ D {\bf 100}, 076018 (2019).

\bibitem{Olugh:2019nej}
  O.~Olugh, Z.~L.~Li and B.~S.~Xie,
  Phys.\ Lett.\ B {\bf 802}, 135259 (2020).
\bibitem{Kohlfurst:2015zxi}
  C.~Kohlf\"urst,
  arXiv:1512.06082 [hep-ph].

\bibitem{Nousch:2015pja}
  T.~Nousch, D.~Seipt, B.~K\"ampfer and A.~I.~Titov,
  Phys.\ Lett.\ B {\bf 755}, 162 (2016).

\bibitem{Aleksandrov:2018zso}
  I.~A.~Aleksandrov, G.~Plunien and V.~M.~Shabaev,
  Phys.\ Rev.\ D {\bf 99}, no. 1, 016020 (2019).

\bibitem{Aleksandrov:2019ddt}
  I.~A.~Aleksandrov and C.~Kohlf\"urst,
  arXiv:1912.09924 [hep-ph].

\bibitem{Schmidt:1998zh}
  S.~M.~Schmidt, D.~Blaschke, G.~Ropke, A.~V.~Prozorkevich, S.~A.~Smolyansky and V.~D.~Toneev,
  Phys.\ Rev.\ D {\bf 59}, 094005 (1999).

\bibitem{Linder:2015vta}
  M.~F.~Linder, C.~Schneider, J.~Sicking, N.~Szpak and R.~Sch\"utzhold,
  Phys.\ Rev.\ D {\bf 92}, no. 8, 085009 (2015).

\end{thebibliography}
\end{document}